\documentclass[12pt,twoside]{article}
\usepackage[mathscr]{eucal}
\usepackage{amsmath,amsfonts,amssymb,amsthm,mathabx,empheq}
\usepackage{times}
\usepackage{pdfsync}
\usepackage{cite}
\usepackage{url}
\usepackage{hyperref}
\usepackage{tensor}
\usepackage{color}
\usepackage{multicol}
\usepackage{bbold}

\advance\textheight\topskip
\allowdisplaybreaks[1]

\newcommand\td{\text{d}}

\newcommand{\p}{\partial}

\newcommand{\be}{\begin{equation}}
\newcommand{\ee}{\end{equation}}
\newcommand{\bea}{\begin{eqnarray}}
\newcommand{\eea}{\end{eqnarray}}

\def\bz{\bar z}

\newcommand*\xbar[1]{%
  \hbox{%
    \vbox{%
      \hrule height 0.5pt % The actual bar
      \kern0.3ex%         % Distance between bar and symbol
      \hbox{%
        \kern-0.0em%      % Shortening on the left side
        \ensuremath{#1}%
        \kern-0.0em%      % Shortening on the right side
      }%
    }%
  }%
}

\DeclareFontFamily{OT1}{rsfs}{} \DeclareFontShape{OT1}{rsfs}{m}{n}{
<-7> rsfs5 <7-10> rsfs7 <10-> rsfs10}{}
\DeclareMathAlphabet{\mycal}{OT1}{rsfs}{m}{n}

\begin{document}
\title{Angular momentum and memory effect}

\author{Pujian Mao, Jun-Bao Wu, and Xiaoning Wu}

\date{}

\def\mytitle{Angular momentum and memory effect}

\begin{flushright}
\tt USTC-ICTS/PCFT-23-03
\end{flushright}

\addtolength{\headsep}{4pt}

\begin{centering}

  \vspace{1cm}

  \textbf{\Large{\mytitle}}

  \vspace{1.5cm}
{\large Pujian Mao$^a$, Jun-Bao Wu$^{a, c}$, and Xiaoning Wu$^{b,d}$ }

\vspace{.5cm}

\begin{minipage}{.9\textwidth}\small \it  \begin{center}
     ${}^{a}$ Center for Joint Quantum Studies and Department of Physics,\\
     School of Science, Tianjin University, 135 Yaguan Road, Tianjin 300350, P. R. China
 \end{center}
\end{minipage}

\vspace{0.3cm}

\begin{minipage}{.9\textwidth}\small \it  \begin{center}
    ${}^{b}$ Institute of Mathematics, Academy of Mathematics and Systems Science\\
    and Hua Loo-Keng Key Laboratory,
    Chinese Academy of Sciences,\\ 55 Zhongguancun East Road, Beijing 100190, P. R. China
 \end{center}
\end{minipage}

\vspace{0.3cm}

\begin{minipage}{.9\textwidth}\small \it  \begin{center}
    ${}^{c}$ Peng Huanwu Center for Fundamental Theory,\\ Hefei, Anhui 230026, P. R. China
 \end{center}
 \end{minipage}

\vspace{0.3cm}

\begin{minipage}{.9\textwidth}\small \it  \begin{center}
    ${}^{d}$ School of Mathematical Sciences, University of Chinese Academy of Sciences,\\ Beijing 100049, P. R. China
 \end{center}

\end{minipage}

\end{centering}

%\begin{center}
%Emails:pjmao@tju.edu.cn%$^{\clubsuit}$$^{\diamondsuit}$
%\end{center}

\vspace{1cm}

\begin{center}
\begin{minipage}{.9\textwidth}
\textsc{Abstract}. It is a long-standing problem in general relativity that the notion of angular momentum of an isolated system has supertranslation ambiguity. In this paper, we argue that the ambiguity is essentially because of the gravitational wave memory. When properly subtracting the memory effect of the observer, one can introduce a supertranslation invariant definition of the angular momentum at null infinity.
 \end{minipage}
\end{center}
\thispagestyle{empty}
\section{Introduction}

More than 60 years ago, Bondi and collaborators established an elegant framework \cite{Bondi:1962px,Sachs:1962wk} of formulating the Einstein equation as a characteristic initial value problem to understand the gravitational
radiation in full Einstein theory. In Bondi's framework, a surprising result is that the asymptotic symmetry group consists of the semidirect product of the group of globally defined conformal transformations of the unit 2-sphere, i.e., the Lorentz group, and an infinite dimensional Abelian normal subgroup, the supertranslations. Consequently, the unexpected enhancement from translations to supertranslations arises crisis in the definition of angular momentum in gravitational system with radiation, the long-standing problem of supertranslation ambiguity issue of angular momentum \cite{Penrose}, see also \cite{Winicour,Prior,Streubel,Held,Ashtekar:1981bq,Geroch:1981ut,Ashtekar,Penrose:1982wp,Dray:1984rfa} for previous efforts and \cite{Szabados:2004xxa} for a recent review. The most direct resolution for this ambiguity issue is simply to modify the known definition of the angular momentum by adding some terms with respect to the transformation law of the supertranslation and to construct a supertranslation invariant definition of angular momentum \cite{Compere:2019gft,Chen:2021szm,Compere:2021inq,Chen:2021kug,Javadinezhad:2022hhl,Javadinezhad:2022ldc}. Nevertheless, the current progress along this direction is mainly about the mathematical construction and the physical meaning of such modifications, in particular, to the observer who detects the angular momentum, is less known.
The aim of the present work is to fill in this gap. To achieve that, we need first to figure out the essence of the supertranslation ambiguity.

In this paper, we argue that the reason of the ambiguity issue of angular momentum is because of the gravitational memory effect \cite{memory,Braginsky:1986ia,1987Natur,Christodoulou:1991cr,Wiseman:1991ss,Thorne:1992sdb,Frauendiener}, rather than the infinite dimensional supertranslation. In four dimensions, gravitational memory is mathematically equivalent to supertranslation \cite{Strominger:2014pwa}, see also \cite{Strominger:2017zoo} for a comprehensive review. Naively, our proposal that attributes the supertranslation ambiguity to the memory effect is just rephrasing the problem in an alternative way. However, there are at least two examples to distinguish those two. The first example is that the asymptotic symmetry group of three dimensional Einstein gravity without cosmological constant also includes a supertranslation part \cite{Ashtekar:1996cd,Barnich:2006av}. However, the definition of angular momentum in three dimensions is free of supertranslation ambiguity \cite{Barnich:2015uva}. One can introduce the notion of intrinsic angular momentum in its rest frame if the supermomentum is brought to a constant \cite{Barnich:2015uva}. Because there is no gravitational wave in three dimensional Einstein gravity. Once the supermomentum is brought to a constant, it will not be changed. The second one is from an electromagnetic analog. If one considers a charged observer at rest in the beginning, the charged observer is forced to accelerate by  electromagnetic wave. Finally the observer will be in another inertial frame with certain velocity after the electromagnetic wave passed by. This is the electromagnetic analog of the gravitational memory effect \cite{Bieri:2013hqa}. Definitely, the angular momentum measured by this observer is problematic in the context of special relativity as the final stage and the initial stage of the charged observer are in two different inertial reference systems. And the transition of the charged observer is not
related to the larger gauge transformation though it is equivalent to the electromagnetic memory \cite{Pasterski:2015zua}. The reason is that the large gauge transformation does not reflect the action of the electromagnetic fields on the charged observer. Nevertheless, this has never been a real problem in defining angular momentum in electromagnetism. Because the definition of conserved quantities is normally in the laboratory frame. The observer in this frame is neutral and does not interact with electromagnetic wave. In contrast, no observer can be free of the gravitational effect. In particular, gravitational wave will induce permanent change to the observer which is the memory effect. Hence the subtlety in defining angular momentum in the gravitational system is similar to the problematic situation of a charged observer in electromagnetic wave. All in all, the long-standing problem in general relativity, the ambiguity in the definition of angular momentum, can be summarized as follows. If one gravitational observer is set to be in the rest frame initially, e.g., in the context of post-Minkowskian approximation, the gravitational memory will finally force this observer out of the rest frame. Consequently, the final measurement of angular momentum and hence the flux of the angular moment will include reference frame effect. Note that the observer with memory effect can properly describe the fundamental laws of physics in the context of general principle of relativity. However, physical quantities are transformed covariantly. Reference frame effect is just encoded in the transformations between different observers, e.g., the definition of angular momentum for different Lorentz frames in special relativity. Hence, the key point to resolve the subtlety in the definition of angular momentum is to subtract the gravitational memory effect. This is precisely what we propose for computing the flux of the angular momentum. And the formula of angular momentum when the gravitational memory effect has been subtracted is supertranslation invariant.
When restricted to a stationary case, our result recovers the supertranslation invariant definitions in \cite{Compere:2019gft,Chen:2021szm,Compere:2021inq,Chen:2021kug,Javadinezhad:2022hhl,Javadinezhad:2022ldc}. Considering the equivalence of supertranslation and memory effect in four dimensional gravitational theory \cite{Strominger:2014pwa}, the subtraction of memory effect is realized by a supertranslation. Thereupon, the observer with the supertranslation invariant definition of angular momentum must undergo a supertranslation at the final stage to subtract the memory effect which is characterized by the additional terms to the angular momentum definition in \cite{Compere:2019gft,Chen:2021szm,Compere:2021inq,Chen:2021kug,Javadinezhad:2022hhl,Javadinezhad:2022ldc}. And this should be the physical meaning of modifying the definition of angular momentum with a supertranslation invariant expression.

The organization of this paper is very simple. In Section \ref{two}, we show that the charged observer has a problem with defining conserved quantities for electromagnetic wave with memory and the problem has no symmetry origin. This is a very simple example of how the memory effect is the essential obstacle for the observer to define conserved quantities. In Section \ref{three}, we give the definition of angular momentum at null infinity which is free of memory effect. Our definition recovers the known supertranslation invariant definition of angular momentum. The last section is devoted to conclusion and discussions for future directions.
%%%%%%%%%%%%%%%%%%%%%%%%%%%%%%%%%%%%%%%%%%
%%%%%%%%%%%%%%%%%%%%%%%%%%%%%%%%%%%%%%%%%%
%%%%%%%%%%%%%%%%%%%%%%%%%%%%%%%%%%%%%%%%%%

\section{Charged observer in electromagnetism}
\label{two}

Gravitational memory effect reflects the interaction of the observer with gravitational wave. The passage of gravitational wave will cause some permanent changes to the observer. Obviously, any change of the observer would affect its measurement. Let us first demonstrate this by a simpler analog, the electromagnetism. Though an electromagnetic analog of gravitational wave memory is known \cite{Bieri:2013hqa}, it is never a problem for defining conserved quantities for electromagnetism. The reason is that there is a good observer that is in the laboratory frame and does not interact with electromagnetic wave. Physical quantities, such as stress tensor and angular momentum density, are well defined in the laboratory frame. For instance, if we start from the Lagrangian of the electromagnetic theory,
\be
\mathcal{L}=-\frac14 F^{\mu\nu}F_{\mu\nu},
 \ee
the symmetric conserved stress tensor is
\be
T_{\mu\nu}=-F_{\mu\rho}F_{\nu}^{\,\,\,\,\rho}-\eta_{\mu\nu}\mathcal{L},
\ee
and the angular momentum density is \cite{Jackson:1998nia}
\be
M^{\mu\nu\lambda}=x^\mu T^{\nu\lambda}-x^\nu T^{\mu\lambda}.
\ee
However, if we consider a charged observer that is initially set to be in the laboratory frame, it will interact with electromagnetic wave. For simplicity, considering a harmonic wave whose vector potential in the Cartesian coordinates is given by
\be
A_y=-\frac{c B_0}{\omega} \sin \omega(t-\frac{x}{c}),\quad
A_t=A_x=A_z=0,
\ee
a charged particle with certain initial conditions will follow the relativistic trajectory \cite{Krueger:1976si}
\be
\begin{split}
x&=\frac{c}{4}\left(\frac{\Omega}{\omega}\right)^2\left(\tau-\frac{\sin2\omega\tau}{2\omega}\right),\\
y&=c\left(\frac{\Omega}{\omega}\right)\frac{1-\cos\omega\tau}{\omega},\\
z&=0,\\
t&=\tau + \frac14 \left(\frac{\Omega}{\omega}\right)^2 \left(\tau-\frac{\sin2\omega\tau}{2\omega}\right),\\
\end{split}
\ee
where $\tau$ is the proper time, $e$ and $m$ are the electric charge and mass of the observer, and $\Omega=\frac{e B_0}{mc}$ is the cyclotron frequency. It is easy to see that the charged observer is not even in an inertial frame along its trajectory. Hence, one has to introduce a fictitious force or fictitious field in the frame of the charged observer to describe physical process. Suppose that the sources of electromagnetic
radiation only exist for a finite time. Finally, the charged observer will be back to an inertial frame. While the charged observer can finally have a velocity with respect to the laboratory frame which is just the electromagnetic memory \cite{Bieri:2013hqa}. Generically, the charged observer at the initial and final stage is in different Lorentz frames. So the flux of the radiated angular momentum evaluated directly by the charged observer is in principle nonphysical and must be different than the observer in the laboratory frame. For the charged observer, the angular momentum at the final stage should be first transformed back to the laboratory frame through a Lorentz transformation plus a displacement of the spatial origin. Then the difference of the transformed angular momentum at the final stage and the initial angular momentum will be just the flux of angular momentum in the laboratory frame. Alternatively, one can consider that the procedure of transforming back to the laboratory frame at the final stage is to subtract the electromagnetic memory effect.

It seems redundant to consider a charged observer in electromagnetism. But a neutral observer does not exist in gravitational theory at all and any observer must interact with gravitational wave, just like a charged observer in electromagnetic wave. Of course, any observer is as good as they should be in the context of the general principle of relativity. However, the principle is for the description of the fundamental laws of physics. For concrete physical quantities such as angular momentum, the measurements do have reference frame dependence. This is relevant to the physical usefulness of the concept of angular momentum. For instance, only the angular momentum of a star with respect to its rest frame is important for the description of stellar structure rather than the orbital angular momentum with respect to, e.g., the Galaxy \cite{Held}. Correspondingly, it can only make sense to consider the flux of the angular momentum when the effects from the reference frame are properly subtracted as we have commented on previously for the charged observer in electromagnetic wave. This is also what we will propose for defining angular momentum for isolated gravitating systems in the presence of gravitational waves.

%%%%%%%%%%%%%%%%%%%%%%%%%%%%%%%%%%%%%%%%%%
%%%%%%%%%%%%%%%%%%%%%%%%%%%%%%%%%%%%%%%%%%
%%%%%%%%%%%%%%%%%%%%%%%%%%%%%%%%%%%%%%%%%%

\section{Supertranslation invariant angular momentum}
\label{three}

In this paper, we will work in the Newman-Penrose (NP) formalism \cite{Newman:1961qr} and use asymptotic analysis in the Newman-Unti (NU) gauge \cite{Newman:1962cia}. The connection of the NU gauge and the Bondi gauge \cite{Bondi:1962px,Sachs:1962wk} can be found, for instance, in \cite{Barnich:2011ty}. We will use the retarded coordinates $(u,r,z,\bz)$, where $A=(z,\bz)$ are the complex stereographic coordinates. The future null infinity is precisely the submanifold $r\to\infty$, with topology $\ensuremath{\mathbb R}\times S^2$. The metric of the celestial sphere in the complex stereographic coordinates is given by
\be
\td s^2=2\gamma_{z\bz}\td z \td \bz=\frac{2 \td z \td \bz}{P^2_s},\quad\quad P_s=\frac{1+z\bz}{\sqrt{2}}.
\ee
The angular momentum current derived from surface charge analysis in NP formalism in natural units $8\pi G=1$ is \cite{Barnich:2019vzx}
\be\label{AM}
J_Y=-\frac{1}{P_s} \left[Y^{\bz}\left(\Psi_1^0 + \sigma_0\eth\xbar\sigma_0 \right) + Y^{z}\left(\xbar\Psi_1^0 +\xbar \sigma_0\xbar\eth\sigma_0 \right)\right],
\ee
where $Y^A$ is a Killing vector of the celestial sphere, i.e., $D_A Y_B + D_B Y_A=0$, and $D_A$ is for the two-dimensional covariant derivative on the celestial sphere. For other notations, we would also refer to \cite{Barnich:2019vzx}. The angular momentum current has certain reference frame dependence which is somewhat similar to the definition of angular momentum in relativistic theory, such as electromagnetism. The issue in gravitational theory with radiation, e.g., in post-Minkowskian approximation, is that even the observer is initially set to be in the rest frame, it will be out of this frame after the gravitational wave passing by \cite{Grishchuk:1989qa,Zhang:2017rno,Zhang:2017jma,Zhang:2018srn,Hamada:2018cjj,Compere:2018ylh,Mao:2018xcw,Flanagan:2018yzh,Mao:2019sph} because of the gravitational memory effect which is similar to the case of a charged observer in electromagnetic wave.

The gravitational memory in NP formalism is the permanent change of the asymptotic shear $\sigma_0$ \cite{Frauendiener}
\be
\eth^2( \Delta_{\pm} \sigma_0)=-\Delta_{\pm} (\Psi_2^0 + \sigma_0 \p_u \xbar\sigma_0) + \int^\infty_{-\infty} \p_u \sigma_0 \p_u \xbar \sigma_0 \td u,
\ee
where $\Delta_{\pm}$ denotes the permanent change in the retarded time. This formula basically is indicating that once there is gravitational radiation, there must be permanent change in the asymptotic shear $\sigma_0$. Then the angular momentum flux computed from the definition \eqref{AM} will inevitably include the effect from gravitational memory which we would refer to as reference frame effect. Hence, it is reasonable to consider that the $\sigma_0$ dependence in the angular momentum current \eqref{AM} is purely reference frame effect that should be subtracted from the definition. The physical angular momentum at null infinity is supposed to be defined in the asymptotically shear-free frame \cite{Held}, i.e., the good cut $\sigma_0=0$, see, e.g., \cite{Adamo:2009vu} for a review on the physically significant effects of asymptotically shear-free null geodesic congruences. Such a configuration is called the canonical gauge in \cite{Veneziano:2022zwh}. One can consider this frame as the asymptotic gravitational laboratory frame though no observer can always stay in this frame due to the gravitational memory effect \cite{Held}. Thus, the strategy for computing the angular momentum flux is as follows. Initially the observer should be chosen as in a good cut. Finally the observer will not be in a good cut due to the memory effect. Nevertheless, one can always use a supertranslation to turn off the asymptotic shear. The angular momentum flux should be computed by the difference between the final data in a good cut and the initial data. The supertranslation at the final stage to turn off the asymptotic shear $\sigma_0$ is precisely the procedure to subtract the reference effect in the definition of angular momentum, see also \cite{Held,Flanagan:2014kfa,Ashtekar:2019rpv} for relevant discussions. Supposing that the gravitational system is back to the stationary case after the gravitational wave passed, the asymptotic shear $\sigma_0$ does not have the retarded time dependence. Under a supertranslation which is characterized by the transformation along the time direction $u'=u+f(z,\bz)$, the asymptotic shear is transformed as $\sigma_0'=\sigma_0+\eth^2 f$. The solution for the good cut is simply a solution of $f$ for the differential equation $\eth^2 f=-\sigma_0$. Note that gravitational memory based upon physically realistic systems is only of an electric type \cite{Winicour:2014ska}, namely only changing the real part of the asymptotic shear. Hence it is always possible to turn off the asymptotic shear arisen by the memory effect, considering that it is initially set to be in a good cut. For a generic case, one has to complexify the time coordinate $u$ to solve the good cut equation \cite{Adamo:2009vu}.

The transformation law of the Weyl tensor component $\Psi_1^0$ under supertranslation is \cite{Barnich:2016lyg}
\be
{\Psi'}_1^0=\Psi_1^0 - 3 \eth f \Psi_2^0.
\ee
Note that we consider the supertranslation at the final stage where the gravitational wave has passed and the spacetime is back to being stationary for which one has $\p_u\sigma_0=\Psi_3^0=\Psi_4^0=0$. Consequently, the angular momentum current is deduced to
\be\label{preliminary}
J_{G}=-\frac{1}{P_s}\left[ Y^{\bz}{\Psi_{G}}_1^0 +Y^z{\xbar{\Psi}_{G}}_1^0\right],
\ee
where we use the subscript $G$ to denote the quantities in the good cut. In the present case, it is just ${\Psi_G}_1^0=\Psi_1^0 - 3 \eth f \Psi_2^0$. This formula of angular momentum is equivalent to the known supertranslation invariant definitions of angular momentum in \cite{Compere:2019gft,Chen:2021szm,Compere:2021inq,Chen:2021kug,Javadinezhad:2022hhl,Javadinezhad:2022ldc}. To manifest the relation, we need to write the NU solution \cite{Newman:1962cia} in the Bondi gauge and to switch to the $(-,+,+,+)$ signature. The metric is given by
\begin{multline}
ds^2=\left[1-\frac{2m}{\rho}+ O(\rho^{-2})\right]du^2 + 2\left(1+O(\rho^{-2})\right)\td u \td \rho \\
- \left[D^z C_{zz}+ \frac{4}{3 \rho} (N_{z}+u \p_{z} m)-\frac{1}{8\rho}\p_z(C_{AB}C^{AB}) + O(\rho^{-2})\right] du dz\\
 - \left[D^{\bz}C_{\bz\bz} + \frac{4}{3 \rho} (N_{\bz}+u \p_{\bz} m) -\frac{1}{8\rho}\p_{\bz}(C_{AB}C^{AB}) + O(\rho^{-2})\right] du d\bz \\
 - \left[C_{zz}  + O(\rho^{-1})\right]dz^2
 - \left[C_{\bz\bz}  + O(\rho^{-1})\right]d\bz^2
 -2\left[\rho^2 \gamma_{z\bz} +  O(1)\right]dzd\bz,
\end{multline}
where the NU radial coordinate $r$ is related to the Bondi radial coordinate $\rho$ as \cite{Barnich:2011ty}
\be
r=\rho+\frac{1}{16}C_{AB}C^{AB} \frac{1}{\rho} + O(\rho^{-3}),
\ee
and the NP quantities are related to the fields in the metric as
\be
\Psi_2^0=-m-\frac{1}{16}\p_u (C_{AB}C^{AB}),\quad \frac{2\sigma_0}{P_s^2}=C_{\bz\bz},\quad C_{z\bz}=0,\quad \frac{\xbar\Psi_1^0}{P_s}=-N_z-u \p_z m.
\ee
Inserting those relations into \eqref{preliminary}, one can obtain the angular momentum current in the Bondi gauge for the stationary case as
\be
J_{G}=Y^A\left[N_A + u \p_A m - 3m \p_A f \right].
\ee
Note that we are now in a good cut, so the time coordinate is $u'$. Hence $\p_A u'=0$. Using integration by part and the fact that $Y^A$ is a Killing vector on the celestial sphere, the angular momentum current is reduced to
\be\begin{split}\label{old}
J_{G}&=Y^A\left[N_A - m \p_A( u+f) - 2 m \p_A f \right]\\
&=Y^A\left[N_A -  m \p_A u' - 2 m \p_A f \right]\\
&=Y^A\left[N_A - 2 m \p_A f \right].
\end{split}\ee
Direct calculation shows that $f=-\frac12 c$ where $c$ is the closed potential of asymptotic shear $C_{AB}$ which is defined in \cite{Chen:2021szm}. Hence, to our surprise, the last expression in \eqref{old} recovers explicitly the supertranslation invariant definition of angular momentum proposed by Chen, Wang, Wang, and Yau \cite{Chen:2021szm}. Note that there is an extra term $-\frac14 Y^A C_{AB} D_E C^{BE}$ in the angular momentum in \cite{Chen:2021szm}. But we have chosen a good cut initially, so this extra term vanishes for the good cut case. At the final stage, the memory effect only changes the electric part of the asymptotic shear. So it can be decomposed as $C_{AB}=-(2D_A D_B - \gamma_{AB}D^2)f$. Then this extra term is a total derivative \cite{Compere:2016jwb,Javadinezhad:2022hhl}
\be
-\frac14 Y^A C_{AB} D_E C^{BE}=\frac14 D^A \left[Y^B C_{AB}(D^2+2)f+\frac12 Y_A \left((D^2+2)f\right)^2\right],
\ee
where one needs to apply the commutator $[D^2,D_A]f=D_A f$ on the celestial sphere and the fact that $Y^A$ is a Killing vector on the celestial sphere, namely, $D_A Y_B=- D_B Y_A$ and $D_A Y^A=0$.
Thus, by subtracting the memory effect and writing the angular momentum in the good cut, we end up with a supertranslation invariant definition of angular momentum. That means the angular momentum expression in the good cut is a supertranslation invariant one. Actually, the key point is to subtract the memory effect and to define the angular momentum in the frame with a fixed asymptotic shear. Thus the angular momentum defined in any fixed cut should be supertranslation invariant by construction. We will postpone commenting on this point and its possible applications in the last section.

To close this section, it is worthwhile to point out that there are ambiguities when deriving the potential of the asymptotic shear. This can be seen clearly from the expansion of the spin weighted spherical harmonics \cite{Newman:1966ub}. More precisely, the $l=0,1$ components in the spherical harmonic expansion of the potential can not be fixed by construction. Those four components correspond to the translation ambiguity. It is obvious that the $l=0$ component would not affect the definition of angular momentum. So this ambiguity is equivalent to the choice of spatial origin for defining angular momentum in special relativity.

%%%%%%%%%%%%%%%%%%%%%%%%%%%%%%%%%%%%%%%%%%
%%%%%%%%%%%%%%%%%%%%%%%%%%%%%%%%%%%%%%%%%%
%%%%%%%%%%%%%%%%%%%%%%%%%%%%%%%%%%%%%%%%%%

\section{Concluding remarks}

In this paper, we argue that the supertranslation ambiguity in the definition of angular momentum for isolated gravitational system is essentially because of the gravitational memory effect. Then we propose a new formula of angular momentum which is free of memory effect, namely the memory effect has been subtracted from the definition of angular momentum. We choose the good cut $\sigma_0=0$ to construct the angular momentum. Nevertheless, one can indeed choose any cut to define the angular momentum. The main point of our proposal is that one should properly subtract the gravitational memory effect and maintain the evaluation of the angular momentum always in the same cut. The physical consequence of other cuts for the definition of angular momentum has been addressed in \cite{Veneziano:2022zwh}. It should be very meaningful to test our proposal in the post-Minkowskian approach which has more direct applications in gravitational wave detection. The radiated angular momentum becomes a relatively urgent issue \cite{Damour:2020tta,Veneziano:2022zwh,Manohar:2022dea,DiVecchia:2022owy,Javadinezhad:2022ldc} since the seminal work \cite{Bern:2019nnu} for high-accuracy calculation of post-Minkowskian dynamics. In particular, setting the asymptotic shear in the intrinsic gauge defined in \cite{Veneziano:2022zwh}, we expect that the angular momentum flux in \cite{Damour:2020tta} can be recovered from the supertranslation invariant definition, for which one needs to apply the connections between the asymptotic expansion in the Bondi framework and the post-Minkowskian expansion \cite{Blanchet:2020ngx}.  Meanwhile, it is important to point out that our prescription is valid in general for defining conserved quantities for observers with memory effect. More precisely, our prescription can be directly applied for defining conserved quantities from the near horizon symmetries \cite{Donnay:2015abr} that are compatible with the black hole memory effect \cite{Donnay:2018ckb,Rahman:2019bmk,Adami:2021nnf}, which has a very important application for the understanding of the black hole soft hairs \cite{Hawking:2016msc}.

%%%%%%%%%%%%%%%%%%%%%%%%%%%%%%%%%%%%%%%%%%
%%%%%%%%%%%%%%%%%%%%%%%%%%%%%%%%%%%%%%%%%%
%%%%%%%%%%%%%%%%%%%%%%%%%%%%%%%%%%%%%%%%%%
\section*{Acknowledgments}

P.M. would like to thank Glenn Barnich, Geoffrey Comp\`{e}re, and Reza Javadinezhad for enlightening discussions and Glenn Barnich again for long term collaborations and support in relevant research topics. This work is supported in part by the National Natural Science Foundation of China under Grants No.~11905156, No.~11975164, No.~11935009, No.~11731001, No.~12275350, No.~12247103, No.~12047502 and Natural Science Foundation of Tianjin under Grant No.~20JCYBJC00910.

%%%%%%%%%%%%%%%%%%%%%%%%%%%%%%%%%%%%%%%%%%
%%%%%%%%%%%%%%%%%%%%%%%%%%%%%%%%%%%%%%%%%%
%%%%%%%%%%%%%%%%%%%%%%%%%%%%%%%%%%%%%%%%%%

%\bibliography{ref}

\begin{thebibliography}{10}

\bibitem{Bondi:1962px}
H.~Bondi, M.~G.~J. van~der Burg, and A.~W.~K. Metzner, ``{Gravitational waves
  in general relativity. 7. Waves from axisymmetric isolated systems},''
  \href{http://dx.doi.org/10.1098/rspa.1962.0161}{{\em Proc. Roy. Soc. Lond.}
  {\bfseries A269} (1962) 21--52}.

\bibitem{Sachs:1962wk}
R.~K. Sachs, ``{Gravitational waves in general relativity. 8. Waves in
  asymptotically flat space-times},''
  \href{http://dx.doi.org/10.1098/rspa.1962.0206}{{\em Proc. Roy. Soc. Lond.}
  {\bfseries A270} (1962) 103--126}.

\bibitem{Penrose}
R.~Penrose, ``{Some unsolved problems in classical general relativity},'' in
  {\em {Seminar on Differential Geometry}}, S.-T. Yau, ed., pp.~631--668.
\newblock Princeton Univ. Press, Princeton, 1982.

\bibitem{Winicour}
J.~Winicour, ``{Some Total Invariants of Asymptotically Flat Space‐Times},''
  \href{http://dx.doi.org/10.1063/1.1664652}{{\em J. Math. Phys.} {\bfseries 9}
  (1968) 861--867}.

\bibitem{Prior}
C.~R. {Prior}, ``{Angular Momentum in General Relativity. I. Definition and
  Asymptotic Behaviour},'' \href{http://dx.doi.org/10.1098/rspa.1977.0073}{{\em
  Proc. Roy. Soc. Lond. A} {\bfseries 354} (1977) 379--405}.

\bibitem{Streubel}
M.~{Streubel}, ``{``Conserved'' quantities for isolated gravitational
  systems},'' \href{http://dx.doi.org/10.1007/BF00759549}{{\em Gen. Rel. Grav.}
  {\bfseries 9} (1978) 551--561}.

\bibitem{Held}
J.~Winicour, ``{Angular momentum in general relativity},'' in {\em {General
  Relativity and Gravitation:One Hundred Years After the Birth of Albert
  Einstein. Volume 2}}, A.~Held, ed., pp.~71--96.
\newblock Plenum Press, New York, 1980.

\bibitem{Ashtekar:1981bq}
A.~Ashtekar and M.~Streubel, ``{Symplectic Geometry of Radiative Modes and
  Conserved Quantities at Null Infinity},''
  \href{http://dx.doi.org/10.1098/rspa.1981.0109}{{\em Proc. Roy. Soc. Lond. A}
  {\bfseries 376} (1981) 585--607}.

\bibitem{Geroch:1981ut}
R.~P. Geroch and J.~Winicour, ``{Linkages in general relativity},''
  \href{http://dx.doi.org/10.1063/1.524987}{{\em J. Math. Phys.} {\bfseries 22}
  (1981) 803--812}.

\bibitem{Ashtekar}
A.~Ashtekar and J.~Winicour, ``{Linkages and Hamiltonians at null infinity},''
  \href{http://dx.doi.org/10.1063/1.525283}{{\em J. Math. Phys.} {\bfseries 23}
  (1982) 2410--2417}.

\bibitem{Penrose:1982wp}
R.~Penrose, ``{Quasilocal mass and angular momentum in general relativity},''
  \href{http://dx.doi.org/10.1098/rspa.1982.0058}{{\em Proc. Roy. Soc. Lond. A}
  {\bfseries 381} (1982) 53--63}.

\bibitem{Dray:1984rfa}
T.~Dray and M.~Streubel, ``{Angular momentum at null infinity},''
  \href{http://dx.doi.org/10.1088/0264-9381/1/1/005}{{\em Class. Quant. Grav.}
  {\bfseries 1} no.~1, (1984) 15--26}.

\bibitem{Szabados:2004xxa}
L.~B. Szabados, ``{Quasi-Local Energy-Momentum and Angular Momentum in GR: A
  Review Article},'' \href{http://dx.doi.org/10.12942/lrr-2004-4}{{\em Living
  Rev. Rel.} {\bfseries 7} (2004) 4}.

\bibitem{Compere:2019gft}
G.~Comp\`ere, R.~Oliveri, and A.~Seraj, ``{The Poincar\'e and BMS flux-balance
  laws with application to binary systems},''
  \href{http://dx.doi.org/10.1007/JHEP10(2020)116}{{\em JHEP} {\bfseries 10}
  (2020) 116}, \href{http://arxiv.org/abs/1912.03164}{{\ttfamily
  arXiv:1912.03164 [gr-qc]}}.

\bibitem{Chen:2021szm}
P.-N. Chen, M.-T. Wang, Y.-K. Wang, and S.-T. Yau, ``{Supertranslation
  invariance of angular momentum},''
  \href{http://dx.doi.org/10.4310/ATMP.2021.v25.n3.a4}{{\em Adv. Theor. Math.
  Phys.} {\bfseries 25} no.~3, (2021) 777--789},
  \href{http://arxiv.org/abs/2102.03235}{{\ttfamily arXiv:2102.03235 [gr-qc]}}.

\bibitem{Compere:2021inq}
G.~Comp\`ere and D.~A. Nichols, ``{Classical and Quantized General-Relativistic
  Angular Momentum},'' \href{http://arxiv.org/abs/2103.17103}{{\ttfamily
  arXiv:2103.17103 [gr-qc]}}.

\bibitem{Chen:2021kug}
P.-N. Chen, M.-T. Wang, Y.-K. Wang, and S.-T. Yau, ``{BMS Charges Without
  Supertranslation Ambiguity},''
  \href{http://dx.doi.org/10.1007/s00220-022-04390-1}{{\em Commun. Math. Phys.}
  {\bfseries 393} no.~3, (2022) 1411--1449},
  \href{http://arxiv.org/abs/2107.05316}{{\ttfamily arXiv:2107.05316 [gr-qc]}}.

\bibitem{Javadinezhad:2022hhl}
R.~Javadinezhad, U.~Kol, and M.~Porrati, ``{Supertranslation-invariant dressed
  Lorentz charges},'' \href{http://dx.doi.org/10.1007/JHEP04(2022)069}{{\em
  JHEP} {\bfseries 04} (2022) 069},
  \href{http://arxiv.org/abs/2202.03442}{{\ttfamily arXiv:2202.03442
  [hep-th]}}.

\bibitem{Javadinezhad:2022ldc}
R.~Javadinezhad and M.~Porrati, ``{Supertranslation-Invariant Formula for the
  Angular Momentum Flux in Gravitational Scattering},''
  \href{http://dx.doi.org/10.1103/PhysRevLett.130.011401}{{\em Phys. Rev.
  Lett.} {\bfseries 130} no.~1, (2023) 011401},
  \href{http://arxiv.org/abs/2211.06538}{{\ttfamily arXiv:2211.06538 [gr-qc]}}.

\bibitem{memory}
Y.~B. Zel'dovich and A.~G. Polnarev, ``{Radiation of gravitational waves by a
  cluster of superdense stars},'' {\em Soviet Astronomy} {\bfseries 18} (Aug.,
  1974) 17.

\bibitem{Braginsky:1986ia}
V.~B. Braginsky and L.~P. Grishchuk, ``{Kinematic Resonance and Memory Effect
  in Free Mass Gravitational Antennas},'' {\em Sov. Phys. JETP} {\bfseries 62}
  (1985) 427--430. [Zh. Eksp. Teor. Fiz.89,744(1985)].

\bibitem{1987Natur}
V.~B. {Braginskii} and K.~S. {Thorne}, ``{Gravitational-wave bursts with memory
  and experimental prospects},'' \href{http://dx.doi.org/10.1038/327123a0}{{\em
  Nature} {\bfseries 327} (May, 1987) 123--125}.

\bibitem{Christodoulou:1991cr}
D.~Christodoulou, ``{Nonlinear nature of gravitation and gravitational wave
  experiments},'' \href{http://dx.doi.org/10.1103/PhysRevLett.67.1486}{{\em
  Phys. Rev. Lett.} {\bfseries 67} (1991) 1486--1489}.

\bibitem{Wiseman:1991ss}
A.~G. Wiseman and C.~M. Will, ``{Christodoulou's nonlinear gravitational wave
  memory: Evaluation in the quadrupole approximation},''
  \href{http://dx.doi.org/10.1103/PhysRevD.44.R2945}{{\em Phys. Rev.}
  {\bfseries D44} no.~10, (1991) R2945--R2949}.

\bibitem{Thorne:1992sdb}
K.~S. Thorne, ``{Gravitational-wave bursts with memory: The Christodoulou
  effect},'' \href{http://dx.doi.org/10.1103/PhysRevD.45.520}{{\em Phys. Rev.}
  {\bfseries D45} no.~2, (1992) 520--524}.

\bibitem{Frauendiener}
J.~Frauendiener, ``{Note on the memory effect},''
  \href{http://dx.doi.org/10.1088/0264-9381/9/6/018}{{\em Class. Quant. Grav.}
  {\bfseries 9} (1992) 1639--1641}.

\bibitem{Strominger:2014pwa}
A.~Strominger and A.~Zhiboedov, ``{Gravitational Memory, BMS Supertranslations
  and Soft Theorems},'' \href{http://dx.doi.org/10.1007/JHEP01(2016)086}{{\em
  JHEP} {\bfseries 01} (2016) 086},
  \href{http://arxiv.org/abs/1411.5745}{{\ttfamily arXiv:1411.5745 [hep-th]}}.

\bibitem{Strominger:2017zoo}
A.~Strominger, ``{Lectures on the Infrared Structure of Gravity and Gauge
  Theory},''
\href{http://arxiv.org/abs/1703.05448}{{\ttfamily arXiv:1703.05448 [hep-th]}}.
%%CITATION = ARXIV:1703.05448;%%.

\bibitem{Ashtekar:1996cd}
A.~Ashtekar, J.~Bicak, and B.~G. Schmidt, ``{Asymptotic structure of symmetry
  reduced general relativity},''
  \href{http://dx.doi.org/10.1103/PhysRevD.55.669}{{\em Phys. Rev. D}
  {\bfseries 55} (1997) 669--686},
  \href{http://arxiv.org/abs/gr-qc/9608042}{{\ttfamily arXiv:gr-qc/9608042}}.

\bibitem{Barnich:2006av}
G.~Barnich and G.~Compere, ``{Classical central extension for asymptotic
  symmetries at null infinity in three spacetime dimensions},''
  \href{http://dx.doi.org/10.1088/0264-9381/24/5/F01}{{\em Class. Quant. Grav.}
  {\bfseries 24} (2007) F15--F23},
  \href{http://arxiv.org/abs/gr-qc/0610130}{{\ttfamily arXiv:gr-qc/0610130}}.

\bibitem{Barnich:2015uva}
G.~Barnich and B.~Oblak, ``{Notes on the BMS group in three dimensions: II.
  Coadjoint representation},''
  \href{http://dx.doi.org/10.1007/JHEP03(2015)033}{{\em JHEP} {\bfseries 03}
  (2015) 033}, \href{http://arxiv.org/abs/1502.00010}{{\ttfamily
  arXiv:1502.00010 [hep-th]}}.

\bibitem{Bieri:2013hqa}
L.~Bieri and D.~Garfinkle, ``{An electromagnetic analogue of gravitational wave
  memory},'' \href{http://dx.doi.org/10.1088/0264-9381/30/19/195009}{{\em
  Class. Quant. Grav.} {\bfseries 30} (2013) 195009},
  \href{http://arxiv.org/abs/1307.5098}{{\ttfamily arXiv:1307.5098 [gr-qc]}}.

\bibitem{Pasterski:2015zua}
S.~Pasterski, ``{Asymptotic Symmetries and Electromagnetic Memory},''
  \href{http://dx.doi.org/10.1007/JHEP09(2017)154}{{\em JHEP} {\bfseries 09}
  (2017) 154}, \href{http://arxiv.org/abs/1505.00716}{{\ttfamily
  arXiv:1505.00716 [hep-th]}}.

\bibitem{Jackson:1998nia}
J.~D. Jackson, {\em {Classical Electrodynamics}}.
\newblock Wiley, 1998.

\bibitem{Krueger:1976si}
J.~Krueger and M.~Bovyn, ``{Relativistic Motion of a Charged Particle in a
  Plane Electromagnetic Wave with Arbitrary Amplitude},''
  \href{http://dx.doi.org/10.1088/0305-4470/9/11/008}{{\em J. Phys. A}
  {\bfseries 9} (1976) 1841--1846}.

\bibitem{Newman:1961qr}
E.~Newman and R.~Penrose, ``{An Approach to gravitational radiation by a method
  of spin coefficients},'' \href{http://dx.doi.org/10.1063/1.1724257}{{\em J.
  Math. Phys.} {\bfseries 3} (1962) 566--578}.

\bibitem{Newman:1962cia}
E.~T. Newman and T.~W.~J. Unti, ``{Behavior of Asymptotically Flat Empty
  Spaces},'' \href{http://dx.doi.org/10.1063/1.1724303}{{\em J. Math. Phys.}
  {\bfseries 3} no.~5, (1962) 891}.

\bibitem{Barnich:2011ty}
G.~Barnich and P.-H. Lambert, ``{A note on the Newman-Unti group},'' {\em Adv.
  Math. Phys.} {\bfseries 2012} (2012) 197385,
  \href{http://arxiv.org/abs/1102.0589}{{\ttfamily arXiv:1102.0589 [gr-qc]}}.

\bibitem{Barnich:2019vzx}
G.~Barnich, P.~Mao, and R.~Ruzziconi, ``{BMS current algebra in the context of
  the Newman-Penrose formalism},''
  \href{http://dx.doi.org/10.1088/1361-6382/ab7c01}{{\em Class. Quant. Grav.}
  {\bfseries 37} no.~9, (2020) 095010},
  \href{http://arxiv.org/abs/1910.14588}{{\ttfamily arXiv:1910.14588 [gr-qc]}}.

\bibitem{Grishchuk:1989qa}
L.~P. Grishchuk and A.~G. Polnarev, ``{Gravitational wave pulses with 'velocity
  coded memory.'},'' {\em Sov. Phys. JETP} {\bfseries 69} (1989) 653--657.

\bibitem{Zhang:2017rno}
P.~M. Zhang, C.~Duval, G.~W. Gibbons, and P.~A. Horvathy, ``{The Memory Effect
  for Plane Gravitational Waves},''
  \href{http://dx.doi.org/10.1016/j.physletb.2017.07.050}{{\em Phys. Lett. B}
  {\bfseries 772} (2017) 743--746},
  \href{http://arxiv.org/abs/1704.05997}{{\ttfamily arXiv:1704.05997 [gr-qc]}}.

\bibitem{Zhang:2017jma}
P.~M. Zhang, C.~Duval, and P.~A. Horvathy, ``{Memory Effect for Impulsive
  Gravitational Waves},''
  \href{http://dx.doi.org/10.1088/1361-6382/aaa987}{{\em Class. Quant. Grav.}
  {\bfseries 35} no.~6, (2018) 065011},
  \href{http://arxiv.org/abs/1709.02299}{{\ttfamily arXiv:1709.02299 [gr-qc]}}.

\bibitem{Zhang:2018srn}
P.~M. Zhang, C.~Duval, G.~W. Gibbons, and P.~A. Horvathy, ``{Velocity Memory
  Effect for Polarized Gravitational Waves},''
  \href{http://dx.doi.org/10.1088/1475-7516/2018/05/030}{{\em JCAP} {\bfseries
  05} (2018) 030}, \href{http://arxiv.org/abs/1802.09061}{{\ttfamily
  arXiv:1802.09061 [gr-qc]}}.

\bibitem{Hamada:2018cjj}
Y.~Hamada and S.~Sugishita, ``{Notes on the gravitational, electromagnetic and
  axion memory effects},''
  \href{http://dx.doi.org/10.1007/JHEP07(2018)017}{{\em JHEP} {\bfseries 07}
  (2018) 017}, \href{http://arxiv.org/abs/1803.00738}{{\ttfamily
  arXiv:1803.00738 [hep-th]}}.

\bibitem{Compere:2018ylh}
G.~Comp\`ere, A.~Fiorucci, and R.~Ruzziconi, ``{Superboost transitions,
  refraction memory and super-Lorentz charge algebra},''
  \href{http://dx.doi.org/10.1007/JHEP11(2018)200}{{\em JHEP} {\bfseries 11}
  (2018) 200}, \href{http://arxiv.org/abs/1810.00377}{{\ttfamily
  arXiv:1810.00377 [hep-th]}}. [Erratum: JHEP 04, 172 (2020)].

\bibitem{Mao:2018xcw}
P.~Mao and X.~Wu, ``{More on gravitational memory},''
  \href{http://dx.doi.org/10.1007/JHEP05(2019)058}{{\em JHEP} {\bfseries 05}
  (2019) 058}, \href{http://arxiv.org/abs/1812.07168}{{\ttfamily
  arXiv:1812.07168 [gr-qc]}}.

\bibitem{Flanagan:2018yzh}
E.~E. Flanagan, A.~M. Grant, A.~I. Harte, and D.~A. Nichols, ``{Persistent
  gravitational wave observables: general framework},''
  \href{http://dx.doi.org/10.1103/PhysRevD.99.084044}{{\em Phys. Rev. D}
  {\bfseries 99} no.~8, (2019) 084044},
  \href{http://arxiv.org/abs/1901.00021}{{\ttfamily arXiv:1901.00021 [gr-qc]}}.

\bibitem{Mao:2019sph}
P.~Mao and W.-D. Tan, ``{Gravitational and electromagnetic memory},''
  \href{http://dx.doi.org/10.1103/PhysRevD.101.124015}{{\em Phys. Rev. D}
  {\bfseries 101} no.~12, (2020) 124015},
  \href{http://arxiv.org/abs/1912.01840}{{\ttfamily arXiv:1912.01840 [gr-qc]}}.

\bibitem{Adamo:2009vu}
T.~M. Adamo, C.~N. Kozameh, and E.~T. Newman, ``{Null Geodesic Congruences,
  Asymptotically Flat Space-Times and Their Physical Interpretation},''
  \href{http://dx.doi.org/10.12942/lrr-2009-6}{{\em Living Rev. Rel.}
  {\bfseries 12} (2009) 6}, \href{http://arxiv.org/abs/0906.2155}{{\ttfamily
  arXiv:0906.2155 [gr-qc]}}.

\bibitem{Veneziano:2022zwh}
G.~Veneziano and G.~A. Vilkovisky, ``{Angular momentum loss in gravitational
  scattering, radiation reaction, and the Bondi gauge ambiguity},''
  \href{http://dx.doi.org/10.1016/j.physletb.2022.137419}{{\em Phys. Lett. B}
  {\bfseries 834} (2022) 137419},
  \href{http://arxiv.org/abs/2201.11607}{{\ttfamily arXiv:2201.11607 [gr-qc]}}.

\bibitem{Flanagan:2014kfa}
E.~E. Flanagan and D.~A. Nichols, ``{Observer dependence of angular momentum in
  general relativity and its relationship to the gravitational-wave memory
  effect},'' \href{http://dx.doi.org/10.1103/PhysRevD.92.084057}{{\em Phys.
  Rev. D} {\bfseries 92} no.~8, (2015) 084057},
  \href{http://arxiv.org/abs/1411.4599}{{\ttfamily arXiv:1411.4599 [gr-qc]}}.
  [Erratum: Phys.Rev.D 93, 049905 (2016)].

\bibitem{Ashtekar:2019rpv}
A.~Ashtekar, T.~De~Lorenzo, and N.~Khera, ``{Compact binary coalescences: The
  subtle issue of angular momentum},''
  \href{http://dx.doi.org/10.1103/PhysRevD.101.044005}{{\em Phys. Rev. D}
  {\bfseries 101} no.~4, (2020) 044005},
  \href{http://arxiv.org/abs/1910.02907}{{\ttfamily arXiv:1910.02907 [gr-qc]}}.

\bibitem{Winicour:2014ska}
J.~Winicour, ``{Global aspects of radiation memory},''
  \href{http://dx.doi.org/10.1088/0264-9381/31/20/205003}{{\em Class. Quant.
  Grav.} {\bfseries 31} (2014) 205003},
  \href{http://arxiv.org/abs/1407.0259}{{\ttfamily arXiv:1407.0259 [gr-qc]}}.

\bibitem{Barnich:2016lyg}
G.~Barnich and C.~Troessaert, ``{Finite BMS transformations},''
  \href{http://dx.doi.org/10.1007/JHEP03(2016)167}{{\em JHEP} {\bfseries 03}
  (2016) 167}, \href{http://arxiv.org/abs/1601.04090}{{\ttfamily
  arXiv:1601.04090 [gr-qc]}}.

\bibitem{Compere:2016jwb}
G.~Comp\`ere and J.~Long, ``{Vacua of the gravitational field},''
  \href{http://dx.doi.org/10.1007/JHEP07(2016)137}{{\em JHEP} {\bfseries 07}
  (2016) 137}, \href{http://arxiv.org/abs/1601.04958}{{\ttfamily
  arXiv:1601.04958 [hep-th]}}.

\bibitem{He:2022idx}
X.~He, X.~Wu, and N.~Xie, ``{On the angular momentum of compact binary
  coalescence},'' \href{http://arxiv.org/abs/2201.12824}{{\ttfamily
  arXiv:2201.12824 [gr-qc]}}.

\bibitem{Newman:1966ub}
E.~T. Newman and R.~Penrose, ``{Note on the Bondi-Metzner-Sachs group},''
  \href{http://dx.doi.org/10.1063/1.1931221}{{\em J. Math. Phys.} {\bfseries 7}
  (1966) 863--870}.

\bibitem{Damour:2020tta}
T.~Damour, ``{Radiative contribution to classical gravitational scattering at
  the third order in $G$},''
  \href{http://dx.doi.org/10.1103/PhysRevD.102.124008}{{\em Phys. Rev. D}
  {\bfseries 102} no.~12, (2020) 124008},
  \href{http://arxiv.org/abs/2010.01641}{{\ttfamily arXiv:2010.01641 [gr-qc]}}.

\bibitem{Manohar:2022dea}
A.~V. Manohar, A.~K. Ridgway, and C.-H. Shen, ``{Radiated Angular Momentum and
  Dissipative Effects in Classical Scattering},''
  \href{http://dx.doi.org/10.1103/PhysRevLett.129.121601}{{\em Phys. Rev.
  Lett.} {\bfseries 129} no.~12, (2022) 121601},
  \href{http://arxiv.org/abs/2203.04283}{{\ttfamily arXiv:2203.04283
  [hep-th]}}.

\bibitem{DiVecchia:2022owy}
P.~Di~Vecchia, C.~Heissenberg, and R.~Russo, ``{Angular momentum of
  zero-frequency gravitons},''
  \href{http://dx.doi.org/10.1007/JHEP08(2022)172}{{\em JHEP} {\bfseries 08}
  (2022) 172}, \href{http://arxiv.org/abs/2203.11915}{{\ttfamily
  arXiv:2203.11915 [hep-th]}}.

\bibitem{Bern:2019nnu}
Z.~Bern, C.~Cheung, R.~Roiban, C.-H. Shen, M.~P. Solon, and M.~Zeng,
  ``{Scattering Amplitudes and the Conservative Hamiltonian for Binary Systems
  at Third Post-Minkowskian Order},''
  \href{http://dx.doi.org/10.1103/PhysRevLett.122.201603}{{\em Phys. Rev.
  Lett.} {\bfseries 122} no.~20, (2019) 201603},
  \href{http://arxiv.org/abs/1901.04424}{{\ttfamily arXiv:1901.04424
  [hep-th]}}.

\bibitem{Blanchet:2020ngx}
L.~Blanchet, G.~Comp\`ere, G.~Faye, R.~Oliveri, and A.~Seraj, ``{Multipole
  expansion of gravitational waves: from harmonic to Bondi coordinates},''
  \href{http://dx.doi.org/10.1007/JHEP02(2021)029}{{\em JHEP} {\bfseries 02}
  (2021) 029}, \href{http://arxiv.org/abs/2011.10000}{{\ttfamily
  arXiv:2011.10000 [gr-qc]}}.

\bibitem{Donnay:2015abr}
L.~Donnay, G.~Giribet, H.~A. Gonzalez, and M.~Pino, ``{Supertranslations and
  Superrotations at the Black Hole Horizon},''
  \href{http://dx.doi.org/10.1103/PhysRevLett.116.091101}{{\em Phys. Rev.
  Lett.} {\bfseries 116} no.~9, (2016) 091101},
  \href{http://arxiv.org/abs/1511.08687}{{\ttfamily arXiv:1511.08687
  [hep-th]}}.

\bibitem{Donnay:2018ckb}
L.~Donnay, G.~Giribet, H.~A. Gonz\'alez, and A.~Puhm, ``{Black hole memory
  effect},'' \href{http://dx.doi.org/10.1103/PhysRevD.98.124016}{{\em Phys.
  Rev. D} {\bfseries 98} no.~12, (2018) 124016},
  \href{http://arxiv.org/abs/1809.07266}{{\ttfamily arXiv:1809.07266
  [hep-th]}}.

\bibitem{Rahman:2019bmk}
A.~A. Rahman and R.~M. Wald, ``{Black Hole Memory},''
  \href{http://dx.doi.org/10.1103/PhysRevD.101.124010}{{\em Phys. Rev. D}
  {\bfseries 101} no.~12, (2020) 124010},
  \href{http://arxiv.org/abs/1912.12806}{{\ttfamily arXiv:1912.12806 [gr-qc]}}.

\bibitem{Adami:2021nnf}
H.~Adami, D.~Grumiller, M.~M. Sheikh-Jabbari, V.~Taghiloo, H.~Yavartanoo, and
  C.~Zwikel, ``{Null boundary phase space: slicings, news \& memory},''
  \href{http://dx.doi.org/10.1007/JHEP11(2021)155}{{\em JHEP} {\bfseries 11}
  (2021) 155}, \href{http://arxiv.org/abs/2110.04218}{{\ttfamily
  arXiv:2110.04218 [hep-th]}}.

\bibitem{Hawking:2016msc}
S.~W. Hawking, M.~J. Perry, and A.~Strominger, ``{Soft Hair on Black Holes},''
  \href{http://dx.doi.org/10.1103/PhysRevLett.116.231301}{{\em Phys. Rev.
  Lett.} {\bfseries 116} no.~23, (2016) 231301},
  \href{http://arxiv.org/abs/1601.00921}{{\ttfamily arXiv:1601.00921
  [hep-th]}}.

\end{thebibliography}

\providecommand{\href}[2]{#2}\begingroup\raggedright\endgroup

\end{document}